# A Case Study of Counting the Number of Unique Users in Linear and Non-Linear Trails - A Multi-Agent System Approach

Tanvir Rahman

May 17, 2023

- Date: Wednesday, May 17, 2023
- Supervisor: Dr. Keith Decker
- Assessor: Dr. R Leila Barmaki

**Department of Computer and Information Sciences**

**University of Delaware**

# Abstract

Parks are an important factor in maintaining or increasing the quality of living for an area. Understanding when, how many, and in what ways people use any given park is very important for taking the necessary steps not just for security measures, but providing the users with facilities like clean washrooms, sitting arrangements, and repairing the paths. Park authorities count the number of daily users by using single entry sensors which add to the total count each time it gets a hit. However, understanding the activities of unique users while within the park is something that is typically not done due to lack of manpower and expense. Today, the wide availability of cheap video cameras and inexpensive on-board, networked processing brings with it the possibility of a deeper understanding of park use by a community. Hence, we propose a multi-agent system approach to analyze the activities of unique users of any park using inexpensive cameras on a distributed system. We have used the Jack A. Markell (JAM) Trail in Wilmington, Delaware, and Hall Trail in Newark, Delaware as a case study for validating our approach. First, we installed a set of video cameras, recorded the video input for a fixed period of time, and then we processed this data autonomously, using existing algorithms, and used that to count the unique users of the park and their activities during the time period. Our approach used different attributes of the users like their speed, direction, activity types, color of their dress, gender etc to identify unique users. The cameras share the attributes while communicating one another, and then the construction of the trails of unique users is done centrally. We validated our results against a human count of unique users, and also developed a simulation to test our approach under varying conditions. Our work has actually set a benchmark for this type of work as it is the first of its kind. We have identified several challenges in this application, and have above a seventy(72) percent success rate for our approach.

# 1 Introduction

Parks are an essential part of any community. They offer many benefits, including recreational activities, socialization opportunities, and environmental benefits. In this article, we will discuss the importance of park user and how better to count them in our park areas.

Counting the number of unique users in a park can be an important step in understanding park usage and planning for its improvement. By knowing how many people use the park, park managers and planners can determine how frequently the park is used and when it is most active. This information can help park managers schedule maintenance and cleaning activities to coincide with times when the park is least used, minimizing disruptions to park-goers.

Additionally, understanding how many unique users visit the park, and what they do can help park planners identify patterns of park usage, including what types of activities are most popular, and what amenities or features are being underutilized. This information can be used to make data-driven decisions about park management and to plan future improvements, such as adding new amenities or features that are more likely to attract visitors.

Finally, tracking park usage and the number of unique users can be important for safety and security reasons. By knowing how many people are using the park at any given time, park managers can ensure that the park is staffed appropriately and that there are enough resources available to keep park-goers safe. In the event of an emergency, this information can also be used to quickly account for all park visitors and ensure that everyone has been safely evacuated.

Currently, parks are using motion sensors to count the number of users. Whenever, the motion sensors detect any motion it counts a person. There are certain disadvantages that comes with it. One, it cannot distinguish between animals and people. Second, it counts every time a person goes through it. Therefore, the count they have perhaps is not count of people, and certainly not unique.

Methods of extracting the videos are not the scope of this paper. Rather, we use the information we get from the extracted videos. We use the generated list of the total users, we took average speed, direction, activity type, gender, color of their dresses and use them to count the number of unique users by already existing systems. As this system is the first of its kind, we had to manually check the unique users first to validate our work. While validating the results, we initially got over 72 percent accuracy which is acceptable. We are deploying a distributed approach of multi agent system to increase the accuracy and real-time analysis. It is also necessary to mention that, we made no contribution to the field of image processing. We use the algorithm proposed by Ribo et al. [11].

In this paper,

- we have formulated a problem definition for this problem.

- Furthermore, we collected real-life data and we have gathered the ground truth, i.e., the unique users park.

- We have formulated an idea which has been done online in simulations for these to implement in the real world scenario.

- We have compared different possible set of attributes to determine the minimum number of attributes necessary for having the unique counts.

- We have made users' trail to know from end to end activity map for any user in a park.

- Also, we found which set of attributes should be used for sensor to sensor communication while identifying the unique users. We also showed that not all the sensors should be active at all times to get desired results.

Some of the challenges that we face during this work include setting up cameras in the parks during collecting real life data. Some of the sensors were stolen, and some were damaged due to weather conditions. We deemed some videos were not usable during human count, and that had a effect on the effectiveness. At first, we worked on the linear trail and communicating with neighboring cameras were easy at that point. However, when we decided to work on non-linear trail, the communication was computationally more demanding than that of linear trail.

The rest of the paper is organized as follows: in Section 2 we survey previous work. Then we cover the problem definition, experiments, our approaches, and discussion. Finally in Section 7, we discussed conclusion and future work.

## 2   Previous Works

In this work, we are trying to show the number of attributes needed for counting the unique users in a park, and making the trails of the users among others. For that, sensors, in this case: cameras, need to communicate with one another using a centralised system while counting the number. While making the trails of the users, sensors need to keep track of incoming messages from their neighbors and have to send messages their neighbors.

In their paper, Decker et. al [2] analyzed this need for communication on meta level. Their work can be viewed from three different points. First, from the practitioners point of view, this work presented a set of design equation that can be used to optimize the performance of Distributed Sensor Network(DSN) or explore the design possibilities while considering a given model of agent cost and the required performance bounds.

Secondly, from the view point of distributed AI community, they have looked upon some of the problems first discussed by pioneers in the field like Durfee, Lesser, and Corkill. In their paper, Durfee et al. [3] said that "Our intent is to show that overly specialized organizational structures allow effective network performance in particular problem-solving situations, but that no such organization is appropriate in all situations." Decker et al. also reached the same same abstract conclusion, however, they were able to precisely show the effect of a organizational structure in an environment. They concluded that it is possible to predict performance as well as explaining organizational structure. Thirdly from the view point of general research community, they were able to present a methodology for giving explanation about designing a system by analysis and simulation. In our work, we did not save the video from the park, we use the information extracted from those videos to count the number of unique users. We, then, showed comparison between different number of set of attributes and the results produced using them.

In their article Tran et al. [7] discuss the use of Wireless Sensor Networks (WSNs) in various industries and their reliance on distributed sensors to monitor environmental factors. The communication technologies utilized in WSNs include WiFi, RF, Bluetooth, and ZigBee, with ZigBee being a preferred choice due to its energy-saving capabilities and long-distance data transmission. The article focuses on the fundamentals of ZigBee network technology and its various communication technologies and applications in WSNs. It also explores different scenarios for mobile agents, including routing protocols in WSNs and presents simulation results demonstrating scalability achieved with ZigBee. Finally, the article concludes with ideas for further ZigBee application development. They also talked about saving energy by not using the sensors all the time. In our work, we implemented energy saving of the sensors by not keeping them active all the time. We also use sensors to extract the attributes of the users, however, they used sensors to monitor environmental factors.

Zhu et al. [13] presents a new approach to enhance the efficiency of data collection from wireless sensor networks. The approach involves the deployment of an unmanned aerial vehicle (UAV) to gather aggregated data from cluster heads, and an unmanned ground vehicle equipped with backup batteries, accompanying the UAV to address energy shortage. The objective is to minimize the mission time for complete data collection, which is framed as a coordinated traveling salesman problem with battery constraints. The solution is obtained using a heuristic path planning algorithm. The results demonstrate the superior performance of the proposed approach over other methods, with the appropriate configuration. In our work, we also gather data from distributed sensors for our work, we did not collect the videos, rather collect the attributes of the observation for our work.

Rodgers et al. [6] discuss the use of wireless sensor networks is becoming more common in various scenarios, including environmental monitoring, security, and military applications. However, the distributed nature and autonomous behavior of these networks pose unique challenges. This article suggests that a new combination of electronic engineering and agent technology is necessary to overcome these challenges. The article provides three examples of successful integration of these two fields. These examples show how this combination can address the need for efficient communication and decentralized algorithms to coordinate the behavior of sensors, facilitate the deployment of sensor agent platforms in the field, and enable the development of intelligent agents capable of autonomous data acquisition, fusion, inference, and prediction.

Hla et al. [4] discussed about a middle agent while talking about multi-agent system solutions for wireless sensor networks. They urged the importance of getting rid of redundant data. The intelligent sensor nodes, data-centric sensor nodes, middle agent or directory facilitator, and mobile agents collectively form the agent community, where each component functions as an autonomous agent. Their architecture has directory facilitator, providers, and requesters. This directory facilitators act as bridges between requesters and providers. Their work has terminal sensors collecting the data, and a centralised agent extracts the information needed from those data and compare them to find the uniqueness. Whether collected data from a sensor will be saved or not depends on the information centralised agent has from other sensors. Like our work, the authors also used heterogeneous sensors for their work as they are placed in different location.

In their paper, Mo et al. [5] describe the use of mobile chargers for wireless charging improves the adaptability of Wireless Rechargeable Sensor Networks (WRSNs), but it also presents difficulties in designing the system. To address these challenges, a new method involving mixed-integer linear programming and a unique decomposition technique is suggested to optimize the scheduling, charging time, and transportation time of multiple mobile chargers. This approach aims to minimize energy consumption while ensuring that all sensors have access to power. This research is discussed in a paper on the coordination of multiple mobile chargers in WRSNs. In our paper, we also worked on how to increase the longevity of the sensor battery life by not having them active all the time, they used batteries for that.

We are using cameras as sensors in our distributed sensor network. These cameras are often placed in inhospitable areas. Therefore sometimes, cameras can be lost. In our experiment, we have lost two cameras as well. Hence, when we place cameras in inhospitable areas we have to make sure that the cameras are energy efficient, and that they do not cost much. This issue has been addressed by Rodgers et al. [6]. They have extended the application of wireless sensor network to security and military scenarios. They stated that sensors in these scenarios ideally should be energy efficient, and they should not be costly. They proposed a decentralised system because they wanted to avoid the single point of failure or bottleneck, the computation power needed for this will be shared, and the solution scales as well as the number of devices in the scenario unlike ours. They have proposed an intermediate agent called 'information agent' that can validate the operational decision taken by autonomous agents. These agents can handle the missing data autonomously. They categorise these networks as pervasive. Their work is primarily in military field, and thus, every missing data is important. Even then, they have to use Gaussian process predictions for several environmental parameters. However, when we are estimating number of people in a public place, e.g., park, we do not necessarily have to be accurate, having the nearest estimation can be viewed as success. Having the success rate more than 70 per cent will give us an estimation that we can work with.

Vinyals et al. illustrated the reasons behind sensor networks being one of the most promising technologies [8]. The reasons behind their proposals are:

- emergence of small and cost efficient sensors based upon microelectromechanicalsystem (MEMS);

- advantages offered by them over other monitoring technologies

- their coverage of different real life applications

They mainly focused on challenges faced on the software front in the multi-agent systems(MAS) fields. They also classified the tasks done by sensor networks into four distinct categories and they are:

- localization

- routing

- information processing

- active sensing

In our work, even though our network has localization, and routing system applied to it; we mainly focused on the active sensing part, and the after the information retrieval, we process the information to count the number of unique users as well as making the trails of the users among others.

Wu et al. [9] discussed about multi-agent system designing for wireless sensory networks for large structure health monitoring. The study of structural health monitoring (SHM), which could improve safety and save maintenance costs for engineering structures, has received a lot of interest. The use of wireless sensor networks (WSN) has been investigated recently in an effort to enhance the capabilities of centralized, cable-based SHM systems. To demonstrate the effectiveness of the multi-agent technology, this study describes a multi-agent design method and system evaluation for wireless sensor network-based structural health monitoring. The distributed wireless sensor network can autonomously assign SHM tasks, self-organize the sensor network, and collect various sensor data with the help of six different agents for SHM applications. The strain gauge and PZT sensors are employed in the evaluation process to track the joint failure of an experimental aluminum plate structure and the change in strain distribution. The assessment system is developed with a dedicated sensor network platform that includes the wireless strain node, wireless PZT node, and wireless USB station. The software architecture for many agents is defined based on the hardware platform. Two common types of structural states are discussed together with the multi-agent monitoring approach and its implementation in the validation job. This study demonstrates the effectiveness of multi-agent technology for WSN applications based on SHM on massive aircraft structures. In our work, we also used multiple sensors to collect information about the park users and then aggregate the information from different sensors to count the number of unique users using their attributes, not the video itself. We also defined our multi-agent system architecture using hardware like them. The purpose of this paper is to demonstrate the effectiveness of multi-agent technology in wireless sensor network-based structural health monitoring through a presentation of a design method and system evaluation.

In their paper, a et al. [1] offers a solution to the decentralized control challenge of tracking a moving multi-target, i.e. having different agents as target, system in a crowded environment using a multi-agent network. The issue is broken down into two related minor issues. Finding a reference density path for the multi-target system that the multi-agent network must follow is the first sub-problem. The answer entails estimating the probability density of the multi-target system as a Gaussian mixture density and displaying it in this way. In the second sub-problem, it is identified which specific inputs will direct the agents to monitor the moving targets without colliding with them. In our work, our observations are moving and we are trying to find their trails and we are trying to determine which among them are unique.

Due to the unpredictable mobility of targets and the constrained coverage range of sensors, it is challenging to achieve maximum target coverage in directional sensor networks (DSNs), mentioned by Xu et al. [10]. In order to solve this issue and avoid missed targets or redundant coverage, sensor coordination is required to ensure optimal target coverage with minimal power usage. They reiterated the fact that structural health monitoring has gained the focus from the research community because it reduces the cost significantly. The Hierarchical Target-oriented Multi-Agent Coordination (HiT-MAC), which separates the problem into two tasks—target assignment by a coordinator and assigned target tracking by executors—is one proposed solution. The coordinator uses global monitoring to allocate targets to each executor, whereas the executor simply keeps track of the targets that are given to it. The HiT-MAC integrates a number of useful techniques, including the self-attention module, marginal contribution approximation for the coordinator, and goal-conditional observation filter for the executor, to increase learning efficiency through reinforcement learning. The HiT-MAC performs better than baseline approaches in terms of coverage rate, learning effectiveness, and scalability, according to empirical findings. Additionally, experiments are used to evaluate the effectiveness of the newly introduced framework components. We also conduct an ablative analysis on the effectiveness of the introduced components in the framework.

Parallel health system monitoring was discussed by Yuan et al. [12]. The fundamental concept behind this new technology is the use of multi-agent technology to handle the entire health monitoring system and smart wireless sensors with on-board microprocessors to establish monitoring sensor networks. First network was designed based on Berkeley Mote Mica wireless sensor platform. In this network two kinds of connected sensors were used: piezoelectric sensors and electric resistance wires.

They defined seven kinds of agents for this health monitoring system. Based on these agents the health monitoring system is proposed. Therefore, a health monitoring system based on Mica wireless sensory platform and multi-agent technology was proposed. In our system, we used sensors as agents collecting the info of the targets. Unlike ours, they used a parallel system for extracting the data.

While some of the works discussed used centralised data collection using UAV [13], rechargeable DSN [5] to save energy, we used sensors that communicate with one another to get activated and therefore, they save energy by not being active. We used the duplicate data at various nodes to make a trail of the users by comparing their attributes collected at different nodes, instead of getting rid of the redundant data [4].

# 3  Problem Definition

The park can be a graph, where different positions of the park are nodes, and with each node there will be a sensor. There can be many starting and ending points for this graph.

Sensors will detect users when they are in cameras' range. It will detect the users' attributes, one of them is speed. Based on the speed, we can categorize the users' activities.

We denote the input of our model as a tuple $(\mathbf{S}, \mathbf{P}, \mathbf{O})$, where

- $\mathbf{S} = \{S_1, S_2, ..., S_k\}$ is a set of sensors
- $\mathbf{P}$ is the park graph
- $\mathbf{O} = \{O_1, O_2, ..., O_n\}$ is a set of observations

Each Sensor, $S_i = \langle g, \rho, E, \mathbf{C} \rangle \in \mathbf{S}$ contains the following, where:

- $g$ is the GPS coordinates (i.e. longitude and latitude)
- $\rho$ is the range such that $S_i$ can observe $U_a$ when distance($S_i$, $U_a$) < $\rho$
- $\mathbf{C}$ is the set of capabilities in which the sensor can observe attributes of a user (e.g. speed, height, direction, energy usage).

$\mathbf{P}$ is a graph where $\mathbf{P} = \{\mathbf{S}, \mathbf{E}\}$

- $\mathbf{S}$ is the set of sensors.
- $E$ is the set of unique edges, i.e., trails between sensors(do not contain any other sensors). $\mathbf{E} = \{E_1, E_2, ..., E_{|E|}\}$ where each $E$:
    - $d$ = distance in meters
    - $exit$ = users will use this point to exit the park.

Observation $\mathbf{O} = \{O_1, O_2, ... O_{|O|}\}$ where each $O$:

- $\mathbf{S}$ = Sensor that produces this $O$
- $id$ = unique id of observation
- $a$ = time of the observation
- $C_{a_i}$ = list of sensed user attributes for this $O$

Each $O$ here will be a row of the table.

We denote the output of our model as a set of unique users $\mathbf{U} = \{U_1, U_2, ..., U_{|U|}\}$, and tracks of these unique users $\mathbf{T} = \{T_1, T_2, ..., T_{|U|}\}$.

A sensor's observations are an assignment of perceived user attributes considering sensor quality. Each sensor will initiate reading when a trail user $U_i$ is in its range $\rho$. It will save the time $U_i$ enters the range and the time s/he departs. It will also save the type of activity based on the trail user's speed along with sensor's $g$, i.e., location. The sensor observation will look be a row in a matrix where the element would be $\langle U_{id}, g_x, g_y, a, C_{a_i} \rangle$.

In the end, we want to determine whether sensor observation is a new person of ID of previously seen person by any sensor(centrally). For that, when a sensor observes any person, we will calculate the distance $d^i_{neighbor}$ between it and neighboring sensors. This sensor will also have the speed of that person $s_u \in C_{u_a}$. Moreover, it will also have the time $a$ when the user $U_a$ enters the sensor. With all these info, we can calculate the time when the user $U_a$ should be in any neighboring sensors. We can user the formula $speed = distance/time$ to calculate the time taken. Then, we compare the time take to $a$ and thus, we will know the time when this $U_a$ should appear in the neighboring sensors. We will try to match all the attributes (e.g. type of activity, height, speed) between these two users $U^{S_i}_a \in S_i$, and $U^{S_{i+1}}_a \in S_{i+1}$, or $U^{S_{i-1}}_a \in S_{i-1}$.

We are taking the color of the dress, activity type, height of trailer, presumed sex of users to distinguish the unique ones. While evaluating we want to minimize the number of falsely created new people, and we want to minimize the number of incorrectly assigned past people. We can minimize the error this way and in doing so, we will have an accurate count of unique trail users. We perform these validation against human count and simulation.

# 4  Data Collection

### Real Life Data

In our experiment, we set up some non-homogeneous cameras in the park. They are not homogeneous because some of them are activated 24/7 and some of them get activated only when there are users.

These sensors are located on the park. For our experiment it was a linear trail at first. We have however extended this work to a non-linear trail as well.

These sensors will produce observations having the users dress color, activity type, apparent gender, timestamp, and age group. These cameras are distributed and they are sharing data centrally while comparing, they are using the information of their neighbours from previous observations.

When any person walks into the range of any sensor we will match it with any qualified previous data from its neighbors. Any data that should be in the range calculating the speed and time spent between these two sensors are qualified observation. After that, we compare two data centrally. If they have the same activity type, same dress color, same sex, same age group, and same speed we categorise them as same person. Therefore the second observation is deemed as duplicate. Otherwise, the observation will be deemed as unique.

The attributes we use for this experiments are the minimum number required for these types of works. If we increase the number of attributes, we will have more distinguishable features to compare, and eventually better results. Without time-stamp, using only site specific data, we were having 51 per cent accurate results. Using time stamp, and thus the speed, this results went up to 78 per cent. Now, we are just using the color of top part of the dress of the observations. If we use the bottom part of the dress color, we will not have the problem where two different observations are detected as one.

### Simulated Data

We also made a simulation with that will produce the similar output. This output has more attributes and we compared the results with different set of attributes, and then decide on the significance of each attribute. We made a side by side comparison for the attributes for the real-life data collected from the park and the simulated data.

Similarly, we had setup a non-linear graph where user trails are non-linear. Collecting the data using the sensors, ie. cameras, are the same. We randomized the starting point of walkers, joggers, bikers in the trails. However, we mostly kept the percent of walkers as 40, and the other two as 30 because we had higher numbers of walkers in the real life data. We also randomized the color of the dresses while populating the simulation so that it represents the real world scenario. There were five different tracks in both of those trails in real life. We kept the track probabilities as random too.

While, calculating sensors share the attributes of the users with their neighbouring sensors. We calculated what are the set attributes we need for identifying the users in real time.

We also compare the power usage of the sensors if they are always on and if they are on when they sense users in their radius. This comparison will give us insights as to how we can increase the longevity of the batteries, and thus, save the expenditure of using sensors while capturing the data.

In the simulation, we used these real life data from Jack A Markell trail and Hall trail from Delaware, and replicated them in a thousand simulations. We used mesa server, and had the picture of the said parks as background. We put sensors and we populated the environment with a hundred people for every simulation.

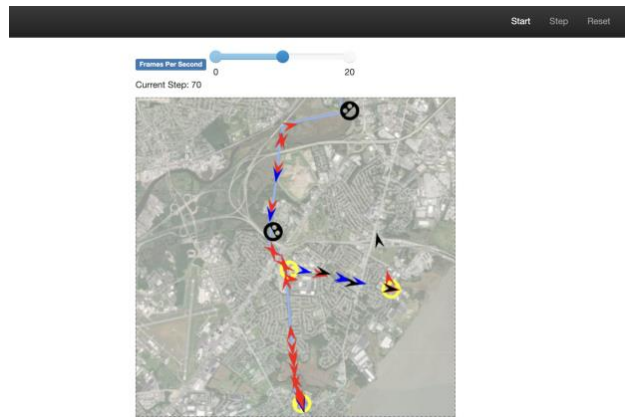

Figure 1: The non-linear trail with active/inactive sensors

In Figure 1, the cameras are yellow when they are capturing the attributes of the observation, and they are black when they are not active because there is no observation in their radius. From this mesa visualization, we get the observation from different sensors in a csv file.

# 5 Our Approaches

### Choosing Attributes

While choosing the attributes that are necessary for identifying the unique users, we needed an algorithm to choose these attributes. First, we calculated the entropy of the attributes using the formula:

$$E(S) = \sum -p_i \log_2(p_i) \qquad (1)$$

Whichever has the lowest entropy were sent to neighboring sensors. For calculation purpose, in our experiment, we do not send all the attributes to the neighboring sensors, rather, we send the important ones chose by this algorithm that will help us distinguish the unique users.

### Identifying Unique Users and Communication between Sensors

One of the attributes that are calculated based on the speed is activity. However, this speed has not been sent to the neighboring sensors. But it was used for two distinguish task-

- determining the activity type
- determining the time of arrival $alpha$ in the next node based on GPS coordinates $g$ and speed from attributes $C$

Now while the attributes of the observation were sent to the neighbor, the estimated arrival time was also sent based on the speed and distance between the two nodes. We used the formula:

$$d = \sqrt{(x_1 - x_2)^2 + (y_1 - y_2)^2} \qquad (2)$$

Here the coordinates $(x_i, y_i)$ are from the GPS $g$ of the sensors. This will not change for any given set environment, or in our case, simulation.

We can send the attributes to all the neighbouring sensors and they would compare the attributes of the coming observation, if any, in the expected arrival time $alpha$. If we have an observation with similar attributes at the estimated arrival time, we conclude that they are the of the same user, and therefore not unique.

### Making a Trail of the Users

When a person is walking towards one camera from another camera, their attributes are being compared by the second camera to decide whether it is the same person or not. The important attributes are being sent to the second camera by the first camera for the comparison.

- there is a set of sensor $< s_1, s_2, ..., s_{|(s)|}>$
- user starts walking on their trail
- trail list, *ListTrails* = *null*
    - user's attributes $C$ are being saved by $s_i$
    - user's attributes $C$ are being sent to $s_{i+1}$
    - user's attributes $C$ are being compared by $s_{i+1}$
    - IF they are not equal
        * append the sensors to the *ListTrails*
    - ELSE
        * continue
    - do loop
- end

When a person comes, ie. an observation, in the range of a sensor it has the attributes of that observation. Using 5.1 that sensor chooses the attributes it wants to send to its neighbors.

### Deciding on Energy Saving Mode

The ideal scenario would be having the sensors active while capturing the attribute of the observations, and all the other time they would not drain they battery energy. In first scenario, we could not do that. We had all the sensors active, all the time. However, that was expensive. We had to think of an alternative where we do not miss any observation.

When a sensor has an observation, it sends the attributes of it to the neighboring sensors, and the estimated time of the arrival. The neighboring sensors get activated during that time, and get the attributes of the observation in their radius. Then they compare the attributes of the observation with those of the first sensor. Then it goes to inactive mode again. That way we save the energy consumed by the sensors. Wait for the communication sent by the neighbors, otherwise be inactive.

- sensors located at the entry and exits are activated all the time , $S_E$
- if $s_i \notin < S_E >$
    - Sensors $< s_2, s_3, ..., s_{|s-1|} >$ are inactive
    - $s_i$ sends the attributes of observation $o_i$ to $s_{neighbors}$ with estimated time based on distance and speed of the observation, and becomes inactive
    - $s_{neighbors}$ gets activated during that time.

- $S_{nieghbors}$ collects the attributes of the observation and compare and make the decision of trails and uniqueness
- $S_{neighbors}$ sends the attributes of observation $o_i$ to its neighbors with estimated time based on distance and speed of the observation, and becomes inactive, if it does not receive another notification of arrival for an observation

# 6  Analysis

We analyzed our work based on two things. One: from the monetary point of view compared to the alternatives, two: what is the efficiency of the results using 1. Ribo et al. [11] showed that the present system costs around ten thousand dollars for setting up the hardware, where our system takes about sixty to hundred dollars per sensor, that means three hundred to six hundred dollars in total and we are set. We will just have to change the batteries once in a while, and even then with our energy saving mode (5.4) we are saving the energy for the batteries as well.

If we can make this system online in the real world, we will not have to save the video anywhere. That will reduce memory-bound dependency and also eliminate the privacy risks. We will only need one real time computation machine that will extract the data from the video feed and erase the videos right after. It is also necessary at this point to mention that the accuracy of the algorithms which we used to extract the attributes from the videos taken by sensors may also played an important role in the efficiency of our work.

As for the linear work we have done, our work produced 70 per cent accuracy with five attributes: Top color, bottom color, activity, age group, and gender. We can see the results for linear work in figure 2 using equation 1. We can see that the color of dress and activity type has played a huge role in deciding the uniqueness of any observation.

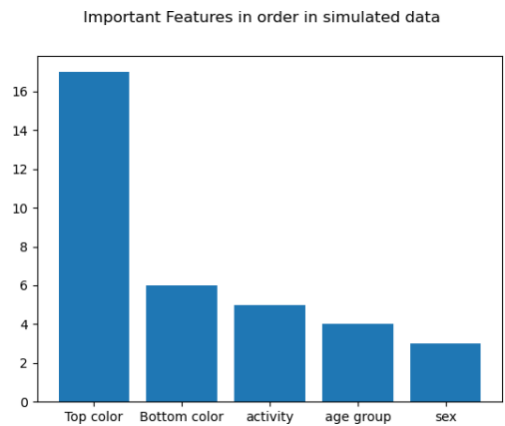

Figure 2: Most important features by the order in linear trail

If we increase more attributes it will increase the efficiency. We will have to keep in mind that, adding more attributes will harm the idea of having the count real time. Because the neighboring sensor will have to compare the attributes with its local attributes before producing into judgement whether or not they are unique observation.

The non-linear data that we have, we used more than the four attributes mentioned. We used bottom color, accessories along with the existing top color, activity, age group, and gender.

We see a change in the results. We got similar 72 per cent accuracy as of the previous 70.

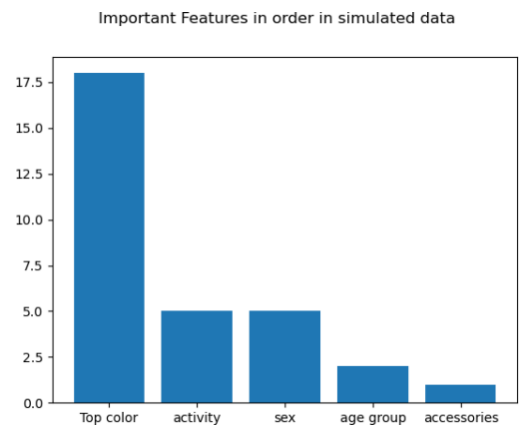

Figure 3: Most important features by the order in non linear trail

Figure 3 shows the important features in order for the non linear data and we can see that the four features we extracted from the linear data are all in top four but one. That actually shows that the features we chose as the attributes for our observations are similar.

Figure 4 shows the power usage by different sensors in the non-linear trail if the sensors are not always on. They only use the power of their batteries when they collect the data of the users. In previous case, when the first sensor starts collecting users' data, all the other sensors are on from that point of time till the end. Therefore they end up consuming more energy than the previous case.

If we see the comparison of these two modes we can actually see how much energy we can save if we do not have sensors that are always on. Originally the power usage of sensors are more as they are always on. Then we implemented the energy saving mode using 5.4 and we have the following results. For this, we use one unit of energy for each time a sensor gets activated. If for an entire

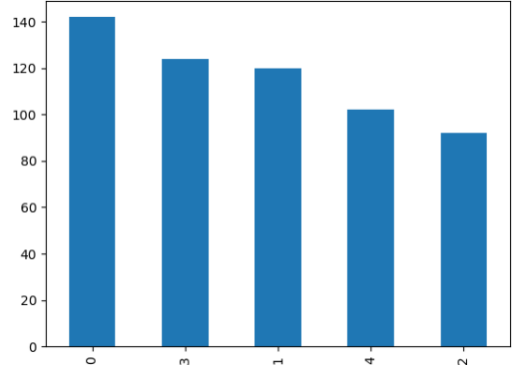

Figure 4: Power usage of sensors not always on non-linear trail

simulation a sensor gets activated for 30 times, then we recon that it has used 30 units of energy.

| mode | linear | non-linear |
| --- | --- | --- |
| usage while energy saving mode on per sensor | 108 | 97 |
| usage while energy saving mode off | 199 | 119 |
| percentage saving by using energy saving mode per sensor | 45 | 18 |

Table 1: Average Energy Usage Per Sensor

We can safely say that implementing the energy saving mode will significantly save the energy of the sensors. It is more effective when the trail is linear as the sensors have less neighbors than a non-linear trail. However, even in non-linear trails the energy saving mode consumes eighty two(82) per cent of the energy it consumes when the mode is not activated leading to save about eighteen(18) per cent of it. The standard deviation of this data set is very low, a mere 1.67.

# 7  Conclusions and Future Work

We have implement our work on two different versions of trails: linear and non-linear trail. However, the data we simulated was the replication based on two different trails, for linear it was Jack A Markell trail, and for non-linear it was Hall trail. Both of them are in Delaware. We made simulations replicating the real life data for our experiments. We validate our work by human count on real life data, and we got similar results on simulated data as well. Our approach has got over seventy two (72) per cent accuracy. Our algorithms was able to save the energy usage of the sensors by at least eighteen(18) per cent.

We want to expand this to other trails in the future. We are making this real-time, however, it is not all in the same place, and more importantly, it is not online for the real life data. We want to do it on an online platform and all in the same place. Here, we are only showing how the energy of the sensors can be used in a better way in terms of usage. The placement of sensors can be optimized to ensure maximum coverage and accuracy while minimizing cost and maintenance requirements. This would require a thorough analysis of the trail environment and visitor behavior patterns. We want to do that in the future. In the future, we can expand the system, incentives could be added to encourage trail usage, such as providing rewards for frequent visitors or those who reach specific distances. This would necessitate the incorporation of a reward system and a user interface to allow visitors to track their progress. Using the real world data, we can compute the confidence interval to estimate the arrival time of the users to the next sensor and therefore better understand accuracy/energy trade-off.

# References

[1] Alaa Z. Abdulghafoor and Efstathios Bakolas. "Motion Coordination of Multi-Agent Networks for Multiple Target Tracking with Guaranteed Collision Avoidance". In: *Journal of Intelligent & Robotic Systems* 107.1 (Jan. 2023). DOI: 10.1007/s10846-022-01786-y. URL: https://doi.org/10.1007/s10846-022-01786-y.

[2] Keith S. Decker and Victor R. Lesser. "Analyzing the Need for Meta-Level Communication". In: 1993.

[3] Edmund H. Durfee, Victor R. Lesser, and Daniel D. Corkill. "Coherent Cooperation Among Communicating Problem Solvers". In: *IEEE Transactions on Computers* C-36.11 (1987), pp. 1275–1291. DOI: 10.1109/TC.1987.5009468.

[4] Khin Haymar Saw Hla, YoungSik Choi, and Jong Sou Park. "The Multi Agent System Solutions for Wireless Sensor Network Applications". In: *Agent and Multi-Agent Systems: Technologies and Applications*. Ed. by Ngoc Thanh Nguyen et al. Berlin, Heidelberg: Springer Berlin Heidelberg, 2008, pp. 454–463. ISBN: 978-3-540-78582-8.

[5] Lei Mo, Angeliki Kritikakou, and Shibo He. "Energy-Aware Multiple Mobile Chargers Coordination for Wireless Rechargeable Sensor Networks". In: *IEEE Internet of Things Journal* 6.5 (2019), pp. 8202–8214. DOI: 10.1109/JIOT.2019.2918837.

[6] Alex Rogers, Daniel D. Corkill, and Nicholas R. Jennings. "Agent Technologies for Sensor Networks". In: *IEEE Intelligent Systems* 24.2 (Mar. 2009), pp. 13–17. URL: https://eprints.soton.ac.uk/267194/.

[7] Hoang Thuan Tran et al. "Mobile agents assisted data collection in wireless sensor networks utilizing ZigBee technology". In: *Bulletin of Electrical Engineering and Informatics* 12.2 (Apr. 2023), pp. 1127–1136. DOI: 10.11591/eei.v12i2.4541. URL: https://doi.org/10.11591/eei.v12i2.4541.

[8] Meritxell Vinyals, Juan A. Rodriguez-Aguilar, and Jesus Cerquides. "A Survey on Sensor Networks from a Multiagent Perspective". In: *The Computer Journal* 54.3 (Feb. 2010), pp. 455–470. ISSN: 0010-4620. DOI: 10.1093/comjnl/bxq018. eprint: https://academic.oup.com/comjnl/article-pdf/54/3/455/1045063/bxq018.pdf. URL: https://doi.org/10.1093/comjnl/bxq018.

[9] Jian Wu et al. "Multi-agent system design and evaluation for collaborative wireless sensor network in large structure health monitoring". In: *Expert Systems with Applications* 37.3 (2010), pp. 2028–2036. ISSN: 0957-4174. DOI: https://doi.org/10.1016/j.eswa.2009.06.098. URL: https://www.sciencedirect.com/science/article/pii/S0957417409006435.

[10] Jing Xu, Fangwei Zhong, and Yizhou Wang. "Learning Multi-Agent Coordination for Enhancing Target Coverage in Directional Sensor Networks". In: *Advances in Neural Information Processing Systems*. Ed. by H. Larochelle et al. Vol. 33. Curran Associates, Inc., 2020, pp. 10053–10064. URL: https://proceedings.neurips.cc/paper_files/paper/2020/file/7250eb93b3c18cc9daa29cf58af7a004-Paper.pdf.

[11] Ribo Yuan et al. *Google Slides: Sign-in*. URL: https://accounts.google.com/v3/signin/identifier?dsh=S1074186085%3A1683820745947300&continue=https%3A%2F%2Fdocs.google.com%2Fpresentation%2Fd%2F1WN5KxlW8W451gc1JATBERSkdLDzgc3tA%2Fedit&followup=https%3A%2F%2Fdocs.google.com%2Fpresentation%2Fd%2F1WN5KxlW8W451gc1JATBERSkdLDzgc3tA%2Fedit&ifkv=Af_xneE4el_Gh7oY4xSBw1_hiDVd_WCS2atSTEAFOy-GqMpzs-_4RxUhgaAjGrKS7pju7r6ZHojIuQ<mpl=slides&osid=1&passive=1209600&service=wise&flowName=GlifWebSignIn&flowEntry=ServiceLogin.

[12] Shenfang Yuan et al. "Distributed structural health monitoring system based on smart wireless sensor and multi-agent technology". In: *Smart Materials and Structures* 15.1 (Dec. 2005), p. 1. DOI: 10.1088/0964-1726/15/1/029. URL: https://dx.doi.org/10.1088/0964-1726/15/1/029.

[13] Yuchao Zhu and Shaowei Wang. "Flying Path Optimization of Rechargeable UAV for Data Collection in Wireless Sensor Networks". In: *IEEE Sensors Letters* 7.2 (2023), pp. 1–4. DOI: 10.1109/LSENS.2023.3237634.